\documentclass[aps,prl,floatfix,preprintnumbers,twocolumn,groupedaddress,nofootinbib,longbibliography]{revtex4-1}

\usepackage{amsmath,amsthm,amssymb,color,psfrag,url,latexsym,graphicx,epstopdf,slashed,xspace,hyperref,enumitem,algorithm,algorithmic}
\usepackage[caption=false]{subfig}

\usepackage[utf8]{inputenc}

\DeclareRobustCommand{\Tab}[1]{Table~\ref{#1}}

\DeclareRobustCommand{\Fig}[1]{Fig.~\ref{#1}}
\DeclareRobustCommand{\Figs}[2]{Figs.~\ref{#1} and \ref{#2}}

\DeclareRobustCommand{\Eq}[1]{Eq.~(\ref{#1})}
\DeclareRobustCommand{\Eqs}[2]{Eqs.~(\ref{#1}) and (\ref{#2})}

\DeclareRobustCommand{\Ref}[1]{Ref.~\cite{#1}}

\definecolor{jdtcolor}{rgb}{0.8,0,0}
\definecolor{emmcolor}{rgb}{0,0.8,0}

\begin{document}
\title{On the Topic of Jets: Disentangling Quarks and Gluons at Colliders}

\author{Eric M. Metodiev}
\email{metodiev@mit.edu}
\affiliation{Center for Theoretical Physics, Massachusetts Institute of Technology, Cambridge, MA 02139, USA}

\author{Jesse Thaler}
\email{jthaler@mit.edu}
\affiliation{Center for Theoretical Physics, Massachusetts Institute of Technology, Cambridge, MA 02139, USA}

\preprint{MIT--CTP 4979}

\begin{abstract}
We introduce jet topics: a framework to identify underlying classes of jets from collider data.
Because of a close mathematical relationship between distributions of observables in jets and emergent themes in sets of documents, we can apply recent techniques in ``topic modeling'' to extract jet topics from data with minimal or no input from simulation or theory. 
As a proof of concept with parton shower samples, we apply jet topics to determine separate quark and gluon jet distributions for constituent multiplicity.
We also determine separate quark and gluon rapidity spectra from a mixed $Z$-plus-jet sample.
While jet topics are defined directly from hadron-level multi-differential cross sections, one can also predict jet topics from first-principles theoretical calculations, with potential implications for how to define quark and gluon jets beyond leading-logarithmic accuracy.
These investigations suggest that jet topics will be useful for extracting underlying jet distributions and fractions in a wide range of contexts at the Large Hadron Collider.
\end{abstract}

\maketitle

% opening
When quarks and gluons are produced in high-energy particle collisions, their fragmentation and hadronization via quantum chromodynamics (QCD) results in collimated sprays of particles called jets.
To extract separate information about quark and gluon jets, though, one typically needs to know the relative fractions of quark and gluon jets in the data sample of interest, estimated by convolving matrix element calculations with non-perturbative parton distribution functions (PDFs).
Recent progress in jet substructure---the detailed study of particle patterns and correlations within jets~\cite{Seymour:1991cb,Seymour:1993mx,Butterworth:2002tt,Butterworth:2007ke,Butterworth:2008iy,Abdesselam:2010pt,Altheimer:2012mn,Altheimer:2013yza,Adams:2015hiv,Larkoski:2017jix}---has offered new avenues to tag and isolate quark and gluon jets~\cite{Nilles:1980ys,Jones:1988ay,Fodor:1989ir,Jones:1990rz,Lonnblad:1990qp,Pumplin:1991kc,Gallicchio:2011xq,Gallicchio:2012ez,Larkoski:2013eya,Larkoski:2014pca,Bhattacherjee:2015psa,FerreiradeLima:2016gcz,Komiske:2016rsd,ATLAS:2016wzt,Davighi:2017hok,Frye:2017yrw}, with recent applications at the Large Hadron Collider (LHC)~\cite{
CMS:2013kfa,Aad:2014gea,Aad:2014bia,Khachatryan:2014dea,Aad:2015owa,Khachatryan:2015bnx,Aad:2016oit,CMS-DP-2016-070,Aaboud:2017jcu}.
Still, there are considerable theoretical uncertainties in the modeling of quark and gluon jets, as well as more fundamental concerns about how to define quark and gluon jets from first principles in QCD~\cite{Banfi:2006hf,Gras:2017jty,Reichelt:2017hts,Mo:2017gzp}.
In particular, quark and gluon partons carry color charge, while jets are composed of color-singlet hadrons, so there is presently no unambiguous definition of ``quark'' and ``gluon'' jet at the hadron level.

% outline
In this letter, we introduce a data-driven technique to extract underlying distributions for different jet types from mixed samples, using quark and gluon jets as an example.
We call our method ``jet topics'' because of a mathematical connection to topic modeling, an unsupervised learning paradigm for discovering emergent themes in a corpus of documents~\cite{blei2012probabilistic}.
Jet topics are defined directly from measured multi-differential cross sections, requiring no inputs from simulation or theory.
In this way, jet topics offer a practical way to define jet classes, allowing us to label ``quark'' and ``gluon'' jet distributions at the hadron level without reference to the underlying partons.

At colliders like the LHC, it is nearly impossible to kinematically isolate pure samples of different jets (i.e.\ quark jets, gluon jets, boosted $W$ jets, etc.).
Instead, collider data consist of statistical mixtures $M_a$ of $K$ different types of jets.
For any jet substructure observable ${\bf x}$, such as jet mass, the distribution $p_{M_a}({\bf x})$ in mixed sample $M_a$ is a mixture of the $K$ underlying jet distributions $p_k({\bf x})$:
\begin{equation}\label{eq:mixtures}
p_{M_a}({\bf x}) = \sum_{k=1}^K f_k^{(a)} \, p_k({\bf x}),
\end{equation}
where $f_k^{(a)}$ is the fraction of jet type $k$ in sample $a$, with $\sum_{k=1}^K f_k^{(a)} = 1$ for all $a$ and $\int d {\bf x} \, p_k({\bf x}) = 1$ for all $k$.

For the specific case of quark ($q$) and gluon ($g$) jet mixtures, we have:
\begin{equation}
\label{eq:mixqg}
p_{M_a}({\bf x}) = f_q^{(a)} \, p_q({\bf x}) + (1 - f_q^{(a)})\, p_g({\bf x}).
\end{equation}
Of course, there are well-known caveats to this picture of jet generation, which go under the name of ``sample dependence''.
For instance, ``quark'' jets from the $Z$+jet process are not exactly identical to ``quark'' jets from the dijet process due to soft color correlations with the entire event~\cite{Gras:2017jty}, though these correlations are power suppressed in the small-jet-radius limit~\cite{Banfi:2010pa,Becher:2016mmh,Kolodrubetz:2016dzb}.
Also, more universal quark/gluon definitions can be obtained using jet grooming methods~\cite{Ellis:2009su,Ellis:2009me,Krohn:2009th,Dasgupta:2013ihk,Larkoski:2014wba,Frye:2016okc,Frye:2016aiz,Marzani:2017mva,Marzani:2017kqd}.
Here, we assume that sample-dependent effects can either be quantified or mitigated, taking \Eq{eq:mixqg} as the starting assumption for our analysis.

Mixed quark/gluon samples were previously studied in the context of Classification Without Labels (CWoLa)~\cite{Metodiev:2017vrx} (see also~\cite{Cranmer:2015bka,blanchard2016classification,Dery:2017fap,Cohen:2017exh,Komiske:2018oaa}).
Via \Eq{eq:mixqg}, one can prove that the optimal binary mixed-sample classifier, $p_{M_1}({\bf x})/p_{M_2}({\bf x})$, is a monotonic rescaling of the optimal quark/gluon classifier, $p_q({\bf x})/p_g({\bf x})$.
This means that a classifier trained to optimally distinguish $M_1$ (e.g.\ $Z$+jet) from $M_2$ (e.g.\ dijets) is optimal for distinguishing quark from gluon jets without requiring jet labels or aggregate class proportions.
The CWoLa framework, though, does not directly yield information about the individual quark and gluon distributions $p_q({\bf x})$ and $p_g({\bf x})$.

With jet topics---and with topic modeling more generally---one can obtain the full distributions $p_k({\bf x})$ and fractions $f_k^{(a)}$ solely from the mixed-sample distributions in \Eq{eq:mixtures}, subject to requirements which will be spelled out below.
As originally posed, topic modeling aims to expose emergent themes in a collection of text documents (a \emph{corpus})~\cite{blei2012probabilistic}.
A \emph{topic} is a distribution over \emph{words} in the \emph{vocabulary}.
\emph{Documents} are taken to be unstructured bags of words.
Each document arises from an unknown mixture of topics: a topic is sampled according to the mixture proportions and then a word is chosen according to that topic's distribution over the vocabulary.
As long as each topic has words unique to it, known as \emph{anchor words}~\cite{arora2012learning,katz2017decontamination}, topic-modeling algorithms can learn the underlying topics and proportions from the corpus alone.

\begin{figure}[t]
\centering
\includegraphics[width=1.0\columnwidth]{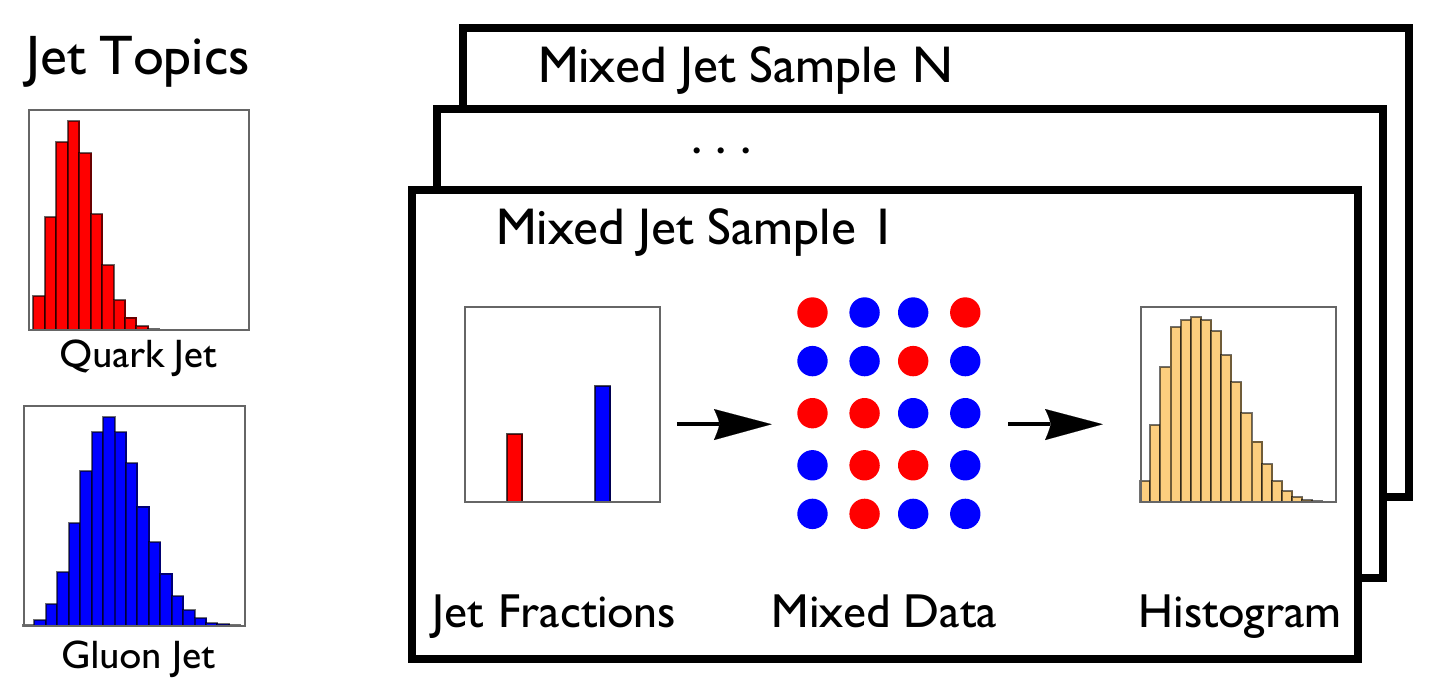}
\caption{The generation of mixed samples of quark and gluon jets, highlighting the correspondence with topic models.
Each jet is either a quark or gluon jet, sampled according to the underlying quark fraction.
The observable is then sampled according to a universal distribution for that jet type.
Each mixed-sample observable distribution is then a mixture of the two universal distributions, giving rise to a ``jet document''.
}
\label{fig:jettopics}
\end{figure}

Intriguingly, the generative process for producing counts of words in a document is mathematically identical to producing jet observable distributions via \Eq{eq:mixtures}, as summarized in \Tab{tab:jettops}.
For the case of quark/gluon jet mixtures, we have suggestively depicted the process of writing ``jet documents'' in \Fig{fig:jettopics}.
Anchor words are analogous to having phase-space regions where each of the underlying distributions is pure, and the presence of these \emph{anchor bins} is necessary for jet topics to yield the underlying ``quark'' and ``gluon'' distributions.

\begin{table}[t]
\begin{tabular}{r@{\hspace{2em}}l}
\hline
\hline
 {\bf Topic Model} & {\bf Jet Distributions}\\
\hline
Word & Histogram bin\\
Vocabulary & Jet observable(s) \\
Anchor word & Pure phase-space region (\emph{anchor bin}) \\
Topic & Type of jet (\emph{jet topic}) \\
Document & Histogram of jet observable(s) \\
Corpus & Collection of histograms \\
\hline
\hline
\end{tabular}
\caption{The correspondence between topic models and jet distributions.  Note that topic modeling treats each document as an unstructured bag of words, in the same way that a collection of jets has no intrinsic ordering.}
\label{tab:jettops}
\end{table}

Due to its theoretical transparency and asymptotic guarantees, we use the \texttt{Demix} method \cite{katz2017decontamination} to extract jet topics, though other algorithms yield comparable results.
The key idea is to undo the mixing of the two fundamental distributions in \Eq{eq:mixtures} by maximally subtracting the two mixtures from one another, such that the zeros of the subtracted distributions correspond to the anchor bins.
Adopting the notation of \Ref{katz2017decontamination}, let $\kappa(M_1|M_2)$ be the largest subtraction amount $\kappa$ such that $p_{M_1}({\bf x}) - \kappa\, p_{M_2}({\bf x})\ge 0$, namely:
\begin{equation}\label{eq:kappa}
\kappa(M_i|M_j) = \min_{\bf x} \frac{p_{M_i}({\bf x})}{p_{M_j}({\bf x})}.
\end{equation}
We refer to $\kappa$ as the \emph{reducibility factor} (equivalently, the minimum of the mixed-sample likelihood ratio).
The \emph{jet topics} $T_1$ and $T_2$ are then the normalized maximal subtractions of $M_2$ from $M_1$,
\begin{equation}
\label{eq:residue}
p_{T_1}({\bf x}) = \frac{p_{M_1}({\bf x}) - \kappa(M_1|M_2) \, p_{M_2}({\bf x})}{1 - \kappa(M_1|M_2)},
\end{equation}
and analogously for $p_{T_2}({\bf x})$.
The jet topics are unique and universal, in that they are independent of the mixtures used to construct them.

The goal is for the topic distributions $p_{T_1}({\bf x})$ and $p_{T_2}({\bf x})$ to match the underlying quark and gluon jet distributions $p_q({\bf x})$ and $p_g({\bf x})$.
There are three required conditions for this to occur. 
Two of them (shared with CWoLa) are \emph{sample independence} and \emph{different purities}, i.e.\ that the jet samples are obtained from \Eq{eq:mixqg} with different values of $f_q^{(a)}$.
The third condition is the presence of anchor bins, which can be stated more formally as:
\begin{itemize}[leftmargin=*]
\item[]\emph{Mutual irreducibility}: Each underlying distribution $p_k({\bf x})$ is not a mixture of the remaining underlying distributions plus another distribution \cite{blanchard2016classification}.
\end{itemize}
Note that this is a much weaker requirement than the distributions being fully separated.
In the quark/gluon context, a necessary and sufficient condition for mutual irreducibility is that the reducibility factors $\kappa(q|g) = \kappa(g|q) = 0$ for feature representation ${\bf x}$.
We later explore the implications of this condition for QCD.
With these three conditions satisfied, the mixture proportions are uniquely determined via the reducibility factors.
Taking $f_q^{(1)}>f_q^{(2)}$, inserting \Eq{eq:mixqg} into \Eq{eq:kappa} yields:
\begin{align}
&\kappa(M_1|M_2) = \frac{1 - f_q^{(1)}}{1 - f_q^{(2)}},  &\kappa(M_2|M_1) =  \frac{f_q^{(2)}}{f_q^{(1)}}.
\end{align}

Even without mutual irreducibility, the extracted jet topics will still relate to the underlying quark and gluon distributions.
Specifically, jet topics yield the ``gluon-subtracted quark distribution'':
\begin{align}
\label{eq:g_sub_q}
p_{q|g}({\bf x}) = \frac{p_q({\bf x}) - \kappa(q|g) \, p_g({\bf x})}{1 - \kappa(q|g)},
\end{align}
and the ``quark-subtracted gluon distribution'', defined analogously.
By universality, the topics calculated from pure samples via \Eq{eq:g_sub_q} and from mixtures via \Eq{eq:residue} are identical.
These may be useful in their own right, particularly if the quark/gluon fractions are uncertain but $\kappa(q|g)$ and $\kappa(g|q)$ can be determined analytically or from simulation (see \Fig{fig:mass} below).

\begin{figure}[t]
\centering
\includegraphics[width=1.0\columnwidth]{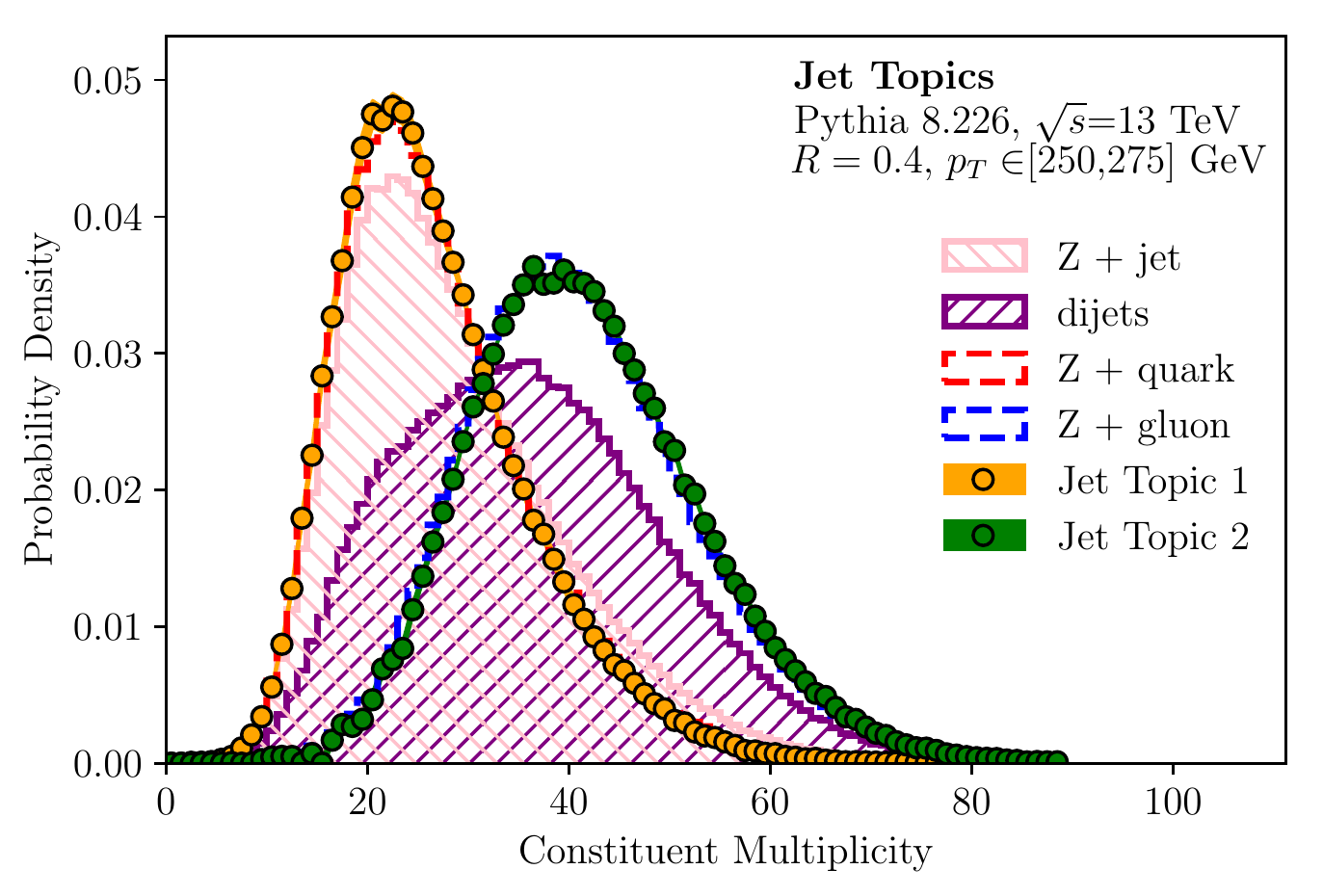}
\caption{The jet topics method applied to constituent multiplicity, starting with $Z$+jet (pink) and dijet (purple) distributions from \textsc{Pythia 8.226}.
There is good agreement between the two extracted jet topics (orange and green) and pure $Z$+quark and $Z$+gluon distributions (red and blue).}
\label{fig:jethists}
\end{figure}

We now turn to a practical demonstration of the jet topics method for realistic quark and gluon samples.
Following \Ref{Gras:2017jty}, we consider two mixed jet processes at the LHC:  the quark-enriched $Z$+jet process and the gluon-enriched dijets process.
See \Ref{Gallicchio:2011xc} for alternative selections for quark- or gluon-enriched samples.
The parton shower \textsc{Pythia 8.226}~\cite{Sjostrand:2006za,Sjostrand:2014zea} is used to generate 500k jets at $\sqrt{s}=13$ TeV including hadronization and multiple parton interactions (i.e.\ underlying event).
Detector-stable, non-neutrino particles are clustered into anti-$k_t$ jets~\cite{Cacciari:2008gp} with radius $R=0.4$ using \textsc{FastJet 3.3.0}~\cite{Cacciari:2011ma}.
The hardest jet(s) in each event (one jet for $Z$+jet and up to two jets for dijets) are selected if they have transverse momentum $p_T \in [250,275]$ GeV and rapidity $|y|\le2$.
These cuts resulted in the $Z$+jet process having (\textsc{Pythia}-labeled) quark fraction $f^{(1)}_q = 0.88$ and the dijet process having $f^{(2)}_q  = 0.37$.
We use the constituent multiplicity within a jet as the feature representation ${\bf x}$, since it is known to be a good quark/gluon discriminant~\cite{Gallicchio:2012ez}.

In \Fig{fig:jethists}, we present the result of extracting two jet topics from these samples.
Shown are the constituent multiplicity distributions from the original $Z$+jet and dijet samples, from \textsc{Pythia}-labeled $Z$+quark and $Z$+gluon samples, and from the jet topics $T_1$ and $T_2$ using \Eq{eq:residue}.
Uncertainties are estimated by assuming $\pm \sqrt{N}$ bin count uncertainties and only considering bins with more than 30 events.
We determine the $\kappa$ values of \Eq{eq:kappa} by selecting the most constraining (anchor) bin: that with the lowest upper uncertainty bar on the ratio.
Remarkably, the two extracted jet topics overlap very well with the underlying quark and gluon distributions, providing practical evidence that \Eq{eq:residue} works as desired, at least for constituent multiplicity.
We verified that similar results could be obtained from samples with different $p_T$ cuts and from mixtures of dijets at different rapidities.
This approach is similar to the template extraction procedure in \Ref{ATLAS:2016wzt}, with the important distinction that the quark/gluon fractions need not be specified a priori.

\begin{figure}[t]
\includegraphics[width=1.0\columnwidth]{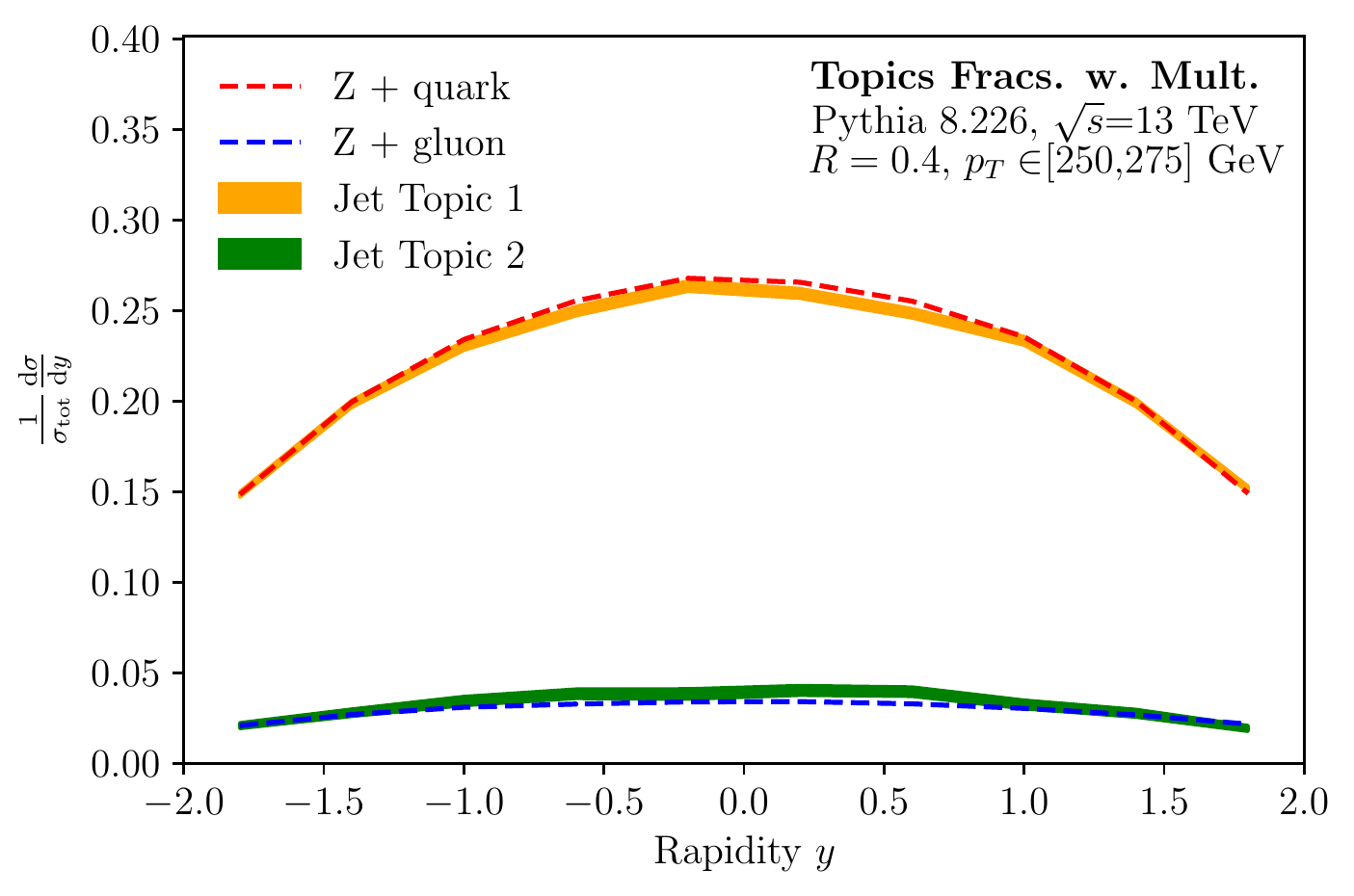}
\caption{Cross sections for jet topics (orange and green) using topic fractions extracted from the $Z$+jet sample across 10 rapidity bins.  The extracted topic cross sections closely track the underlying $Z$+quark and $Z$+gluon cross sections (red and blue).}
\label{fig:cs}
\end{figure}

In \Fig{fig:cs}, we use the extracted jet topics to construct separate jet rapidity spectra for quark and gluon jets in the $Z$+jet samples.
Binning the $Z$+jet sample into 10 rapidity bins in $|y|<2$, we find the mixture of the two topics extracted above that most closely matches the constituent multiplicity histogram in each rapidity bin, minimizing the squared error to find the best mixture.
This is an example of the general problem of extracting sample fractions $f_k^{(a)}$ from various mixed samples.
As desired, the extracted topic cross sections in \Fig{fig:cs} track the true quark and gluon rapidity cross sections.

Thus, just from a collection of mixed-sample histograms, one can make progress toward extracting both the underlying distributions $p_k({\bf x})$ and the fraction of each jet topic $f_k^{(a)}$.
Crucially, \Figs{fig:jethists}{fig:cs} are just novel projections of the hadron-level multi-differential jet cross section $\text{d}^3 \sigma / \text{d} p_T \,  \text{d} y \, \text{d} n_{\rm const}$ on two independent samples, making jet topics implementable on existing LHC jet measurements (e.g.~\cite{Aad:2016oit}).
The agreement between the operationally-defined jet topics and the theoretically-ambiguous quark and gluon distributions may even suggest using mutual irreducibility of the final-state distributions to define ``quark" and ``gluon" jets.

From the perspective of first-principles QCD, the implications of mutual irreducibility are simple yet profound.
For the reducibility factors $\kappa(q|g)$ and $\kappa(g|q)$ to be zero, there must be phase-space regions almost entirely dominated by quark or gluon jets.
In the leading-logarithmic (LL) limit, mutual irreducibility can be achieved with any jet substructure observable that counts the number of parton emissions, such as ``soft drop multiplicity'' \cite{Frye:2017yrw}.
At LL order, quark and gluon jets have the same emission profile, differing only by a color factor in their emission density, $C_F = 4/3$ for quarks and $C_A = 3$ for gluons.
Ignoring the $\Lambda_{\rm QCD}$ regulator, counting these (infinitely many) emissions results in arbitrarily well-separated quark and gluon Poissonian distributions~\cite{Frye:2017yrw}, and therefore mutual irreducibility.
Beyond LL order, though, naive quark/gluon definitions may not lead to mutual irreducibility, since running-coupling, higher-order, and non-perturbative effects generically contaminate the anchor bins.
That said, as long as these effects maintain sample independence (perhaps achieved via grooming), then one can still use \Eq{eq:g_sub_q} to define subtracted ``quark'' and ``gluon'' labels.

Interestingly, many jet substructure observables do not lead to quark/gluon mutual irreducibility, even at LL accuracy.
Consider for instance the jet mass $m$ (or any jet angularity~\cite{Berger:2003iw,Almeida:2008yp,Ellis:2010rwa}).
Jet mass exhibits Casimir scaling at LL order, meaning that the cumulative density functions $\Sigma_i(m)$ are related to each other by $\Sigma_g = \Sigma_q^{C_A/C_F}$~\cite{Larkoski:2013eya,Larkoski:2014pca}.
The probability distributions are then given by $p_i = \text{d} \Sigma_i / \text{d} m$.
Substituting this into \Eq{eq:kappa} immediately yields, for all observables with Casimir scaling:
\begin{align}
\label{eq:kgq}\kappa(g|q) &= \frac{C_A}{C_F}\, \min \Sigma_q^{\frac{C_A}{C_F} - 1} = 0,\\
\label{eq:kqg}\kappa(q|g) &= \frac{C_F}{C_A}\, \min \Sigma_q^{1 - \frac{C_A}{C_F}} = \frac{C_F}{C_A},
\end{align}
since $C_A/C_F = 9/4 > 1$ and $\Sigma$ takes all values between 0 and 1.
Because of \Eq{eq:kqg}, jet mass alone is not sufficient to extract the quark distribution at LL order without additional information.

\begin{figure}[t]
\centering
\includegraphics[width=1.0\columnwidth]{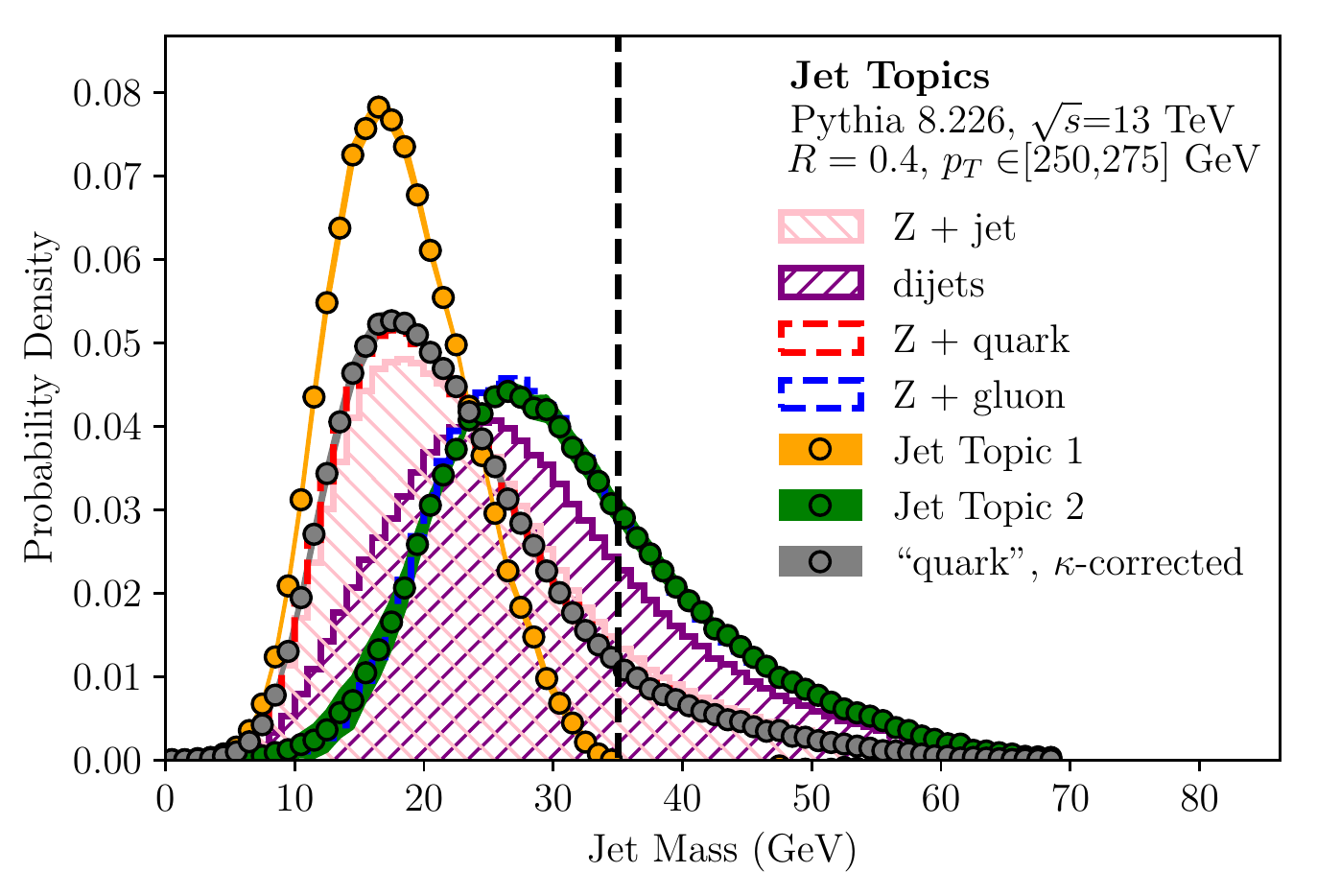}
\caption{The jet topics method applied to jet mass up to 35~GeV (up to the black, dashed line).  The gray curve is the corrected quark topic using \textsc{Pythia} to determine $\kappa(q|g)$, extrapolated beyond 35 GeV by letting jet topic 1 go negative.  There is good agreement between the $\kappa$-corrected quark topic (gray) and the pure $Z$+quark distribution (red).}
\label{fig:mass}
\end{figure}

On the other hand, if the reducibility factors are known, then the subtracted distributions in \Eq{eq:g_sub_q} can be inverted.
This is shown in \Fig{fig:mass} for the jet mass, where the ``quark'' topic has been corrected using the value $\kappa(q|g) = 0.40$ at 35 GeV determined from the \textsc{Pythia} $Z+q/g$ distributions, which is known to differ from the LL expectation~\cite{Gras:2017jty}.
This analysis is performed up to 35 GeV to avoid sample-dependent effects in the high-mass tails of the distributions.
The qualitative behavior of the topics agrees with the LL predictions of \Eqs{eq:kgq}{eq:kqg}: no correction is needed to obtain the ``gluon'' topic, and the ``quark'' topic is a non-trivial mixture of the jet topics.
Given the good agreement seen here, it would be interesting to apply jet topics to groomed jet mass measurements~\cite{Aaboud:2017qwh,CMS-PAS-SMP-16-010}, where grooming is an essential ingredient that allows $\kappa(q|g)$ to be calculated to high precision~\cite{Frye:2016okc,Frye:2016aiz,Marzani:2017mva,Marzani:2017kqd}.

% conclusions
There are many potential uses for the jet topics framework at the LHC.
Focusing just on quark and gluon jets, one often wants to separately measure quark and gluon distributions from mixed data samples, without relying on theory or simulation for fraction estimates.
To determine PDFs, it would be beneficial to isolate different partonic subprocesses, and this could be feasible as long as jet topics is applied both to data and to fixed-order QCD calculations.
Similar subprocess isolation might be useful in mono-jet searches for dark matter by aiding in signal/background discrimination or in setting improved limits on specific new physics models~\cite{Aaboud:2017phn,Sirunyan:2017jix}.
For extracting the strong coupling constant $\alpha_s$ from (groomed) jet shape distributions, it would be beneficial to determine the quark and gluon jet fractions using data-driven methods, since there are uncertainties associated with whether $\alpha_s$ comes multiplied by $C_F$ or $C_A$~\cite{Bendavid:2018nar}.
The extracted topic fractions could be also be used to augment training with CWoLa, since the classifier operating points could then be determined entirely from data.
In heavy ion collisions, quarks and gluons are expected to be modified differently in medium due to their different color charges, and jet topics may allow for fully data-driven studies of separate quark and gluon jet modifications.

In conclusion, phrasing jet mixtures as a topic modeling problem makes available a variety of new and more sophisticated statistical and mathematical tools for jet physics (see e.g.~\cite{blei2003latent,lee1999learning,donoho2004does,arora2012learning,naik2011overview,blanchard2014decontamination,jain2016nonparametric,katz2016mutual,katz2017decontamination,comon2010handbook,reju2009algorithm,sanderson2014class,scott2015rate,ramaswamy2016mixture,ding2015necessary,ding2015learning,arora2013practical,katz2017decontamination}), including recent efforts to determine the appropriate number of topics to use from data~\cite{greene2014many,wang2014topic,zhao2015heuristic}.
We emphasize that jet topics can be applied to any set of multi-differential cross sections---in experiment or in theory---as long as the criteria of sample independence, different purities, and mutual irreducibility are met.
Furthermore, mutual irreducibility need not be assumed if the subtracted distributions in \Eq{eq:g_sub_q} are sufficient for the intended application, or if the reducibility factors are known from theory or simulation.
Of course, experimental studies are needed to understand the systematic and statistical uncertainties associated with jet topics for LHC measurements and searches, and theoretical studies are needed to determine the interplay of jet topics with precision calculations.
It would also be interesting to design jet substructure observables specifically targeted for mutual irreducibility.
More generally, topic models may find applications in collider physics beyond jets and in other disciplines beyond collider physics, since extracting signal and background distributions from mixtures is a ubiquitous challenge faced when analyzing and interpreting rich data sets.

% acknowledgements
\begin{acknowledgments}
The authors would like to thank Mario Campanelli, Timothy Cohen, Philip Harris, Patrick Komiske, Andrew Larkoski, Ian Moult, Benjamin Nachman, Gavin Salam, Clayton Scott, and Wouter Waalewijn for illuminating discussions.
The authors are grateful to Patrick Komiske for generating the jet samples.
This work was supported by the Office of Nuclear Physics of the U.S. Department of Energy (DOE) under grant DE-SC0011090 and the DOE Office of High Energy Physics under grant DE-SC0012567.
Computations in this paper were run on the Odyssey cluster supported by the FAS Division of Science, Research Computing Group at Harvard University.
Cloud computing resources were provided through a Microsoft Azure for Research award.
\end{acknowledgments}

%\newpage

\bibliography{jettopics}

\end{document}